\documentclass{article}
\usepackage[latin9]{inputenc}
\usepackage{geometry}
\usepackage{color}
\usepackage{amsmath}
\usepackage{amssymb}
\usepackage{graphicx}
\usepackage{esint}
\usepackage[numbers]{natbib}

\makeatletter

\usepackage{amsfonts}
\providecommand{\U}[1]{\protect\rule{.1in}{.1in}}

\makeatother

\begin{document}

\title{Quantification and prediction of extreme events in a one-dimensional
nonlinear dispersive wave model}

\author{Will Cousins and Themistoklis P. Sapsis%
\thanks{Corresponding author: sapsis@mit.edu, Tel: (617) 324-7508, Fax: (617)
253-8689%
}\\
 Department of Mechanical Engineering, Massachusetts Institute of
Technology,\\
 77 Massachusetts Av., Cambridge, MA 02139}

\date{\today}

\maketitle
 
\begin{abstract}
The aim of this work is the quantification and prediction of rare
events characterized by extreme intensity in nonlinear waves with
broad spectra. We consider a one-dimensional nonlinear model with
deep-water waves dispersion relation, the Majda-McLaughlin-Tabak (MMT)
model, in a dynamical regime that is characterized by broadband spectrum
and strong nonlinear energy transfers during the development of intermittent
events with finite-lifetime. To understand the energy transfers that
occur during the development of an extreme event we perform a spatially
localized analysis of the energy distribution along different wavenumbers
by means of the Gabor transform. A stochastic analysis of the Gabor
coefficients reveals i) the low-dimensionality of the intermittent
structures, ii) the interplay between non-Gaussian statistical properties
and nonlinear energy transfers between modes, as well as iii) the
critical scales (or critical Gabor coefficients) where a critical
amount of energy can trigger the formation of an extreme event. We
analyze the unstable character of these special localized modes directly
through the system equation and show that these intermittent events
are due to the interplay of the system nonlinearity, the wave dispersion,
and the wave dissipation which mimics wave breaking. These localized
instabilities are triggered by random localizations of energy in space,
created by the dispersive propagation of low-amplitude waves with
random phase. Based on these properties, we design low-dimensional
functionals of these Gabor coefficients that allow for the prediction
of the extreme event well before the nonlinear interactions begin
to occur. 
\end{abstract}

\section{Introduction\label{sec:introduction}}

Extreme or rare events have attracted substantial attention in various
scientific fields both because of their catastrophic impact but also
because of the serious lack of specialized mathematical tools for
the analysis of the underlying physics. Important examples can be
found in i) the environmental field: rogue waves in the ocean \citep{Dysthe08,Akhmediev,muller,xiao13},
 extreme weather and climate events \citep{Neelin,majda10PNAS},
and ii) the engineering field: overloads and failures in power grids
\citep{Kishore12,Pourbeik}, stability loss and capsizing of ships
in mild waves \citep{Kreuzer}. For all of the above applications
it has now been well established that extreme events occur much more
frequently than it was initially believed and that their traditional
characterization as `rare events' (especially in a Gaussian context
where a rare event has practically zero probability) severely underestimates
the frequency of their occurrence. Therefore, it is important to study
them more thoroughly and develop effective algorithms for their prediction.

Extreme events refer to system responses with magnitude that is much
larger than the typical deviation that characterizes the system response.
Thus, from the very nature of these events it can be concluded that
traditional analysis tools restricted to second order statistics would
not be sufficient for their understanding. Apart from their intermittent
properties, another manifestation of the non-Gaussian character of
extreme events is the strong localization of energy in (physical or
modal) space -- a situation that is inherently connected with non-linear
dynamics and transient or persistent instabilities, which has been
shown (see e.g. \citep{sapsis11a,sapsis_majda_mqg}) to be an important
factor that can lead to non-Gaussian statistics.

These characteristics also define the modeling challenges for the
study of these systems with the most important being the interplay
of a few intermittent modes with a large number of modes that act
as `reservoir' of energy for the former. This large set of modes is
usually characterized by a broadband spectrum consisting of dispersive
waves with weakly non-Gaussian statistics that propagate and sporadically
give rise to extreme, localized events. In contrast to this large
set of waves, extreme events are characterized by strong nonlinear
energy transfers and non--Gaussian statistics. Therefore, we have
on the one hand a nearly Gaussian `heat bath' of waves that propagate
in the presence of dispersion which leads to
energy localization in random scales and places, and on the other hand
a nonlinear mechanism that uses the former as excitation to generate
extreme events \citep{osborne}.

It is clear from the above discussion that a mathematical framework
able to handle problems characterized by extreme
events should include higher order statistics and also should be able
to deal with the inherent nonlinear character of the underlying dynamics.
However, the computational cost associated with these requirements
would be enormous since i) the number of physical degrees of freedom
is usually very large and ii) because the description of non-Gaussian
properties and in particular the description of rare events that `live'
in the tails of the distribution requires a substantial amount of
realizations which is very hard to obtain and process in a direct
Monte-Carlo framework. In addition, a purely statistical understanding
cannot provide a rigorous analysis of the underlying physical
mechanisms.

On the other hand, order-reduction approaches based, for example, on
Polynomial Chaos expansions or Proper Orthogonal decompositions have
proven to be of limited applicability in nonlinear systems with intermittency
\citep{majda_branicki_DCDS}. Due to their localized spatial and temporal
character, extreme events carry only small amounts of energy compared
with other global modes that characterize the full response field.
Therefore, standard order-reduction techniques will most likely miss
the essential parts of the extreme event dynamics.

To simulate the dynamical mechanisms that lead to
the generation of extreme events, we use the MMT model, a one-dimensional
nonlinear dispersive equation originally proposed by Majda, McLaughlin,
and Tabak to assess the validity of weak turbulence theory \citep{majda1997}.
MMT admits four-wave resonant interactions and, when coupled with
large scale forcing and small scale damping, admits a rich family
of spectra exhibiting direct and inverse cascades \citep{cai1999,cai2001}.
Zakharov et. al. have also analyzed the MMT model in detail and have
used large amplitude coherent structures present in MMT as models
of extreme ocean waves \citep{zakharov2001,zakharov2004,pushkarev2013}.
In this work, we analyze in detail the `solitonic' coherent structures
in the focusing MMT, which have also been investigated by Cai et.
al. \citep{cai2001}. In their early stages, these localized structures
resemble self-similar spatial collapses and rapidly transfer energy
to small scales where it is dissipated \citep{cai2001}. We are particularly
interested in these localized structures as they generate states which
are extreme compared to the benign background out of which they arise.

In the present work, we first aim to develop analytical and numerical
tools in order to understand how these localized extreme events are
triggered by spatially localized perturbations in the MMT model. We
illustrate that there is a critical spatial lengthscale and a critical
amount of energy associated with it that leads to the occurrence of
extreme solutions. This critical scale is the result of the interplay
between wave dispersion, wave nonlinearity and selective dissipation
that occurs in high wavenumbers. For perturbations of a zero background
state we are able to analyze this phenomena directly by deriving a
family of scale invariant solutions. However, the critical amount
of energy depends also on the background energy level of the system,
the effects of which we analyze numerically. In contrast to the standard
linearized analysis, which considers small Fourier mode perturbations
about about a given state, the framework presented here
considers spatially localized perturbations that are not necessarily
small.

We illustrate that these extreme events are characterized by low-dimensionality
and we use a spatially localized basis, a Gabor basis, with localization
characteristics tuned according to the results of the previous conclusions.
Using the projected information of the extreme events to this localized
basis we perform a statistical analysis of the Gabor coefficients
to reveal the strongly non-Gaussian character associated with the
strongly nonlinear interactions of these modes during an extreme event.
Note that this statistical structure, which is directly connected
to the nonlinear energy transfers that take place, is otherwise `buried'
in the broad-band spectrum of the full wave field and its only signature
in the stochastic field response is the heavy tail statistics.

Finally, we formulate predictive functionals that efficiently characterize
the domain of attraction to the extreme event solutions. These predictive
functionals are formulated in a probabilistic fashion in terms of
the Gabor coefficients that correspond to the critical lengthscales.
Given the current information of the wavefield, they provide the probability
of occurrence of an extreme event in a later time instant. Note that
the propagation of waves (having random phases) in the presence of
dispersion creates conditions for localization of energy in arbitrary
scales and positions in space. The formulated probabilistic functionals
assess these random localizations of energy and quantify the
probability that they will lead to an occurrence of an extreme event
in the future.

\section{A one-dimensional, dispersive nonlinear prototype model with intermittent
events}

We consider the following one-dimensional partial differential equation
originally proposed by Majda, McLaughlin, and Tabak \citep{majda1997}
for the study of 1D wave turbulence: 
\begin{equation}
iu_{t}=\left\vert \partial_{x}\right\vert ^{\alpha}u+\lambda\left\vert \partial_{x}\right\vert ^{-\beta/4}\left(\left\vert \left\vert \partial_{x}\right\vert ^{-\beta/4}u\right\vert ^{2}\left\vert \partial_{x}\right\vert ^{-\beta/4}u\right)+iDu\label{eq:MMT}
\end{equation}
where $u$ is a complex scalar. On the real line, the pseudodifferential
operator $|\partial_{x}|^{\alpha}$ is defined through the Fourier
transform as follows: 
\[
\widehat{\left\vert \partial_{x}\right\vert ^{\alpha}u}\left(k\right)=|k|^{\alpha}\widehat{u}\left(k\right).
\]
This operator may also be defined analogously on a periodic domain.
The MMT equation was introduced on the basis of a simple enough model
to test thoroughly the predictions of weak turbulence theory. In the
context of dispersive nonlinear waves it provides a prototype system
with non-trivial energy transfers between modes or scales, non-Gaussian
statistics with heavy tails, and intermittent events with high intensity,
while remaining accessible to high resolution simulations \citep{majda1997,cai1999,cai2001,grooms12}.
Therefore, it is an ideal basis to assess the performance of probabilistic
quantification algorithms for the occurrence and prediction of extreme
events. Even though the MMT model was originally derived through a
heuristic approach, it was later shown that it can be rigorously obtained
as an approximation of the fully nonlinear wave system equations \citep{trulsen2000}.

In the present work the parameter $\alpha$ is set to 1/2 as this
matches the dispersion relation for deep water waves $\omega^{2}=|k|$.
Setting $\alpha=2$ and $\beta=0$ in \ref{eq:MMT} yields the nonlinear
Schrödinger equation (ignoring the dissipation term). As in \citep{majda1997}
we include dissipation at small scales (modeling e.g. wave breaking
in the context of water waves) through a selective Laplacian operator
$Du,$ defined in Fourier space: 
\[
\widehat{Du}(k)=\left\{ \begin{array}{cl}
-(|k|-k^{\ast})^{2}\hat{u}(k) & |k|>k^{\ast}\\
0 & |k|\leq k^{\ast}
\end{array}\right.
\]
Similar dissipation models have been used in more realistic settings
involving ocean water waves \citep{komen94}. The critical wavenumber
is taken as $k^{\ast}=500$ which is a value that is large enough
so that it allows for the development of nonlinear instabilities that
lead to extreme waves and small enough to create energy cascades in
higher wavenumbers and thus, allow for these waves to exist only for
finite-time.\textbf{\textcolor{red}{{} }}

We choose $\lambda=-4$, which corresponds to the focusing case and
gives rise to four-wave resonant interactions \citep{majda1997} which
are relevant for ocean gravity waves. The latter cannot resonate in
order lower than four if we exclude the short wave gravity-capillary
region of the spectrum where three-wave interactions can occur \citep{komen94}.
Note that even though a connection with the fully nonlinear wave system
would require $\lambda>0$ (defocusing case), we have found through
direct numerical simulations that for the one-dimensional case considered
here such choice results in Gaussian statistics without intermittent
events. This is not the case for the two-dimensional system where
the de-focusing case may lead to non-Gaussian statistics \citep{onorato02}.
On the other hand, for the one-dimensional case the focusing case
results in the development of extreme events (through four-wave resonance)
and to this end such a choice of parameters is suitable for a prototype
system that generates rare events of extreme intensity.

\begin{figure}[ptb]
\centering \includegraphics[width=4.6259in,height=2.0263in]{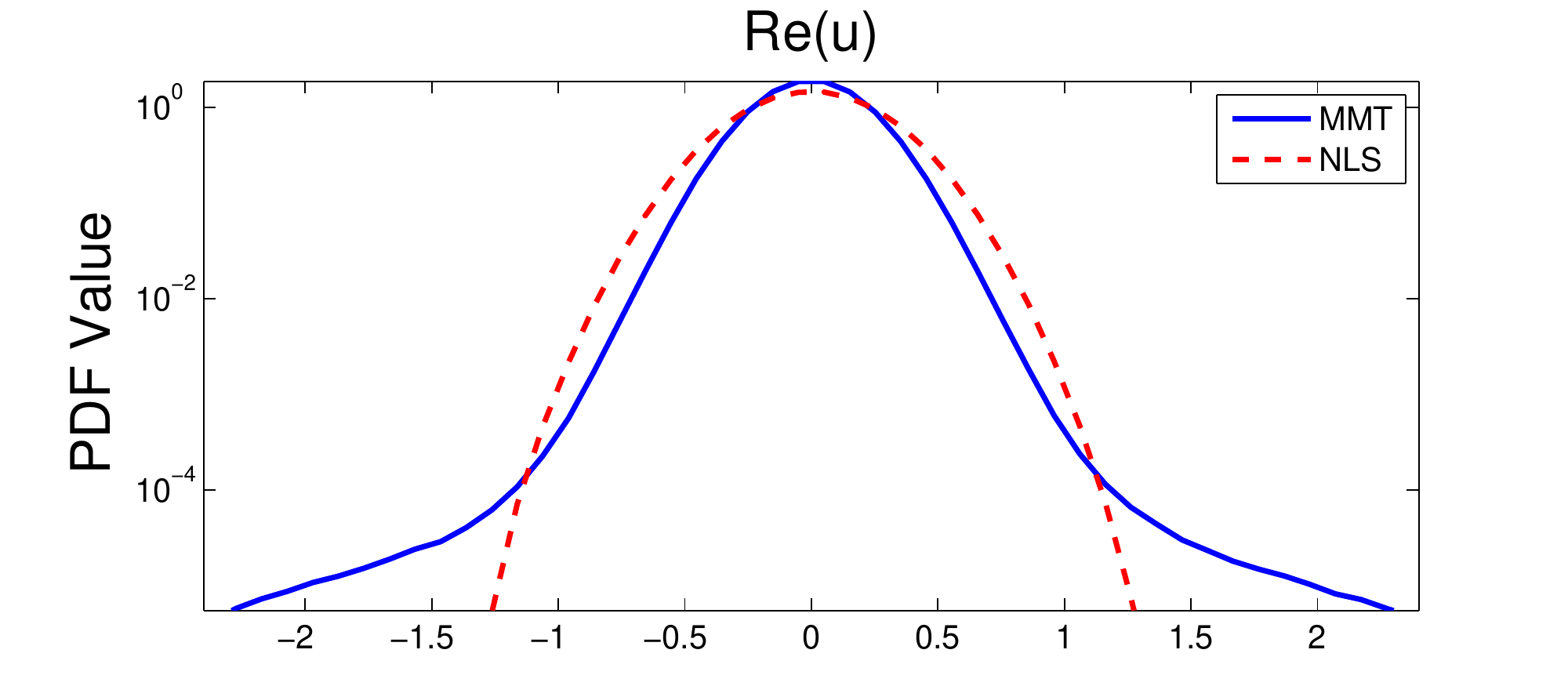}\caption{Probability density for the real part of $u$ for simulations of NLS
and MMT ($\alpha=1/2,\beta=0$)}

\label{fig:PDF_MMT} 
\end{figure}

We consider the evolution of a sum of complex exponentials with independent,
uniformly distributed random phases, meaning that initially $u$ has
a nearly Gaussian distribution. For a linear model, the distribution
would remain nearly Gaussian as the modes evolve independently. Interestingly,
even in simulations of the focusing nonlinear Schrödinger equation
we find that $u$ remains nearly Gaussian. However, for the MMT model
we find that the distribution $u$ develops heavy tails with a power
law decay rate (see Figure~\ref{fig:PDF_MMT}). 

\begin{figure}[ptb]
\centering \includegraphics[width=4.9813in,height=3.1125in]{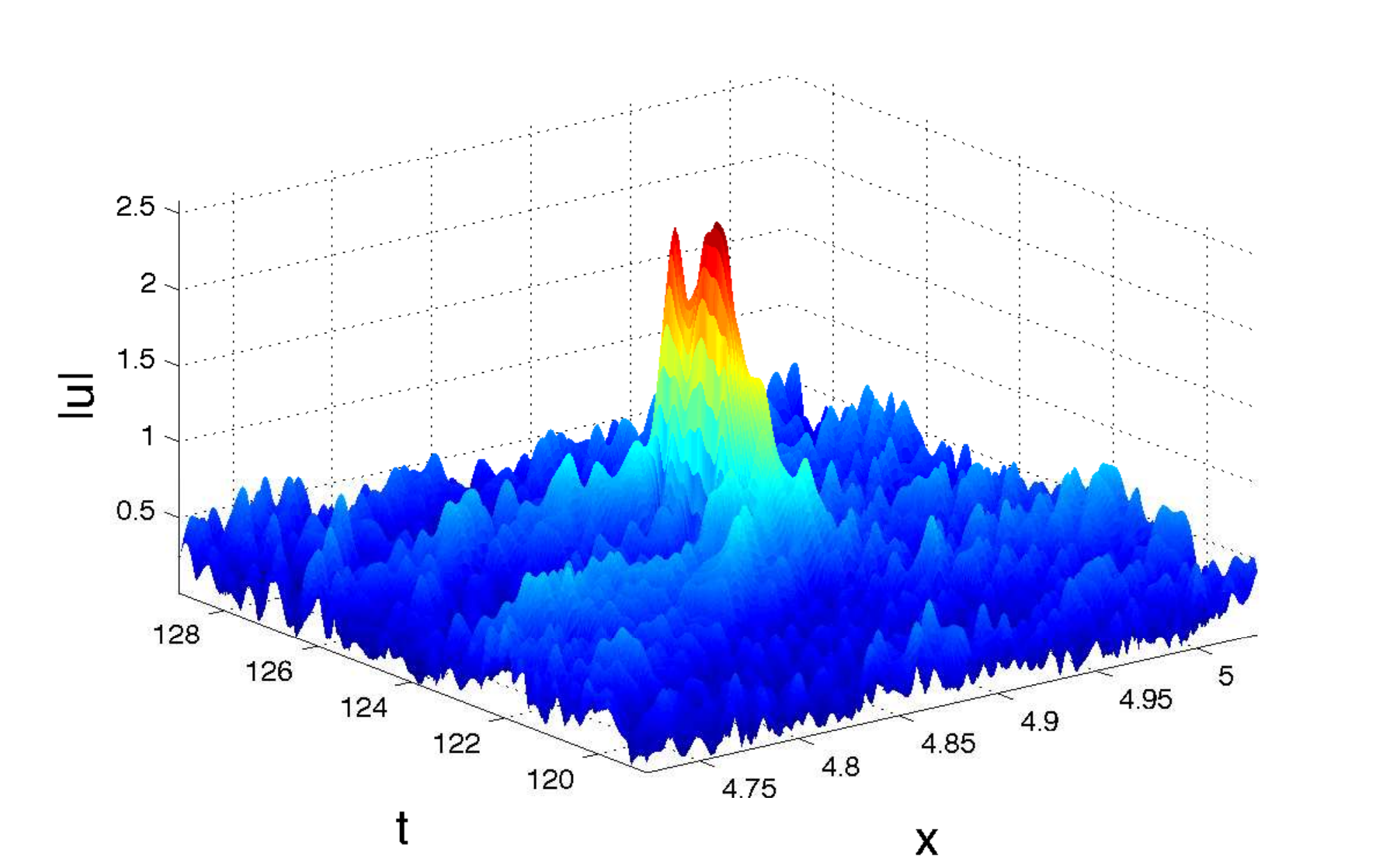}\caption{Example of an extreme event arising out of a weakly non-Gaussian `heat
bath' of dispersive waves with random phase.}

\label{fig:exampleEE} 
\end{figure}

The heavy tails in solutions of MMT are induced by the intermittent
formation and subsequent collapse of localized extreme events arising
out of a nearly Gaussian background. Figure~\ref{fig:exampleEE}
displays the origination and disappearance of such an extreme event.
In their early stages these extreme events resemble the collapses
that are present in focusing MMT with no dissipation. In these collapses,
which have been described by Cai et. al. \citep{cai2001}, energy
is dramatically transferred to smaller scales and the solution experiences
a singularity in finite time. In our simulations, the small scale
dissipation included in (\ref{eq:MMT}) (modeling wave breaking in
the context of water waves) ensures that $u$ remains regular for
all times. Collapse dynamics have been found to induce heavy tailed
statistics in other situations as well, such as the damped-driven
quintic 1D nonlinear Schrödinger equation \citep{chung2011}. 

\subsection{Numerical simulation and computation of statistics\label{sec:simulation_statistics}}

We solve (\ref{eq:MMT}) for $x\in[0,2\pi]$ with periodic boundary
conditions using a Fourier method in space combined with a 4th order
Runge-Kutta exponential time differencing scheme \citep{cox2002,grooms12}.
This scheme requires evaluation of the function $\phi(z)=(e^{z}-1)/z$.
Naive computation of $\phi$ can suffer from numerical cancellation
error for small $z$ \citep{kassam2005}. We use a Padé approximation
code from the EXPINT software package, which does not suffer from
such errors \citep{berland2007}. We use $2^{13}$ Fourier modes with
a time step of $10^{-3}$; results in this work were insensitive to
further refinement in grid size.

\begin{figure}[h]
\centering \includegraphics[width=0.8\textwidth]{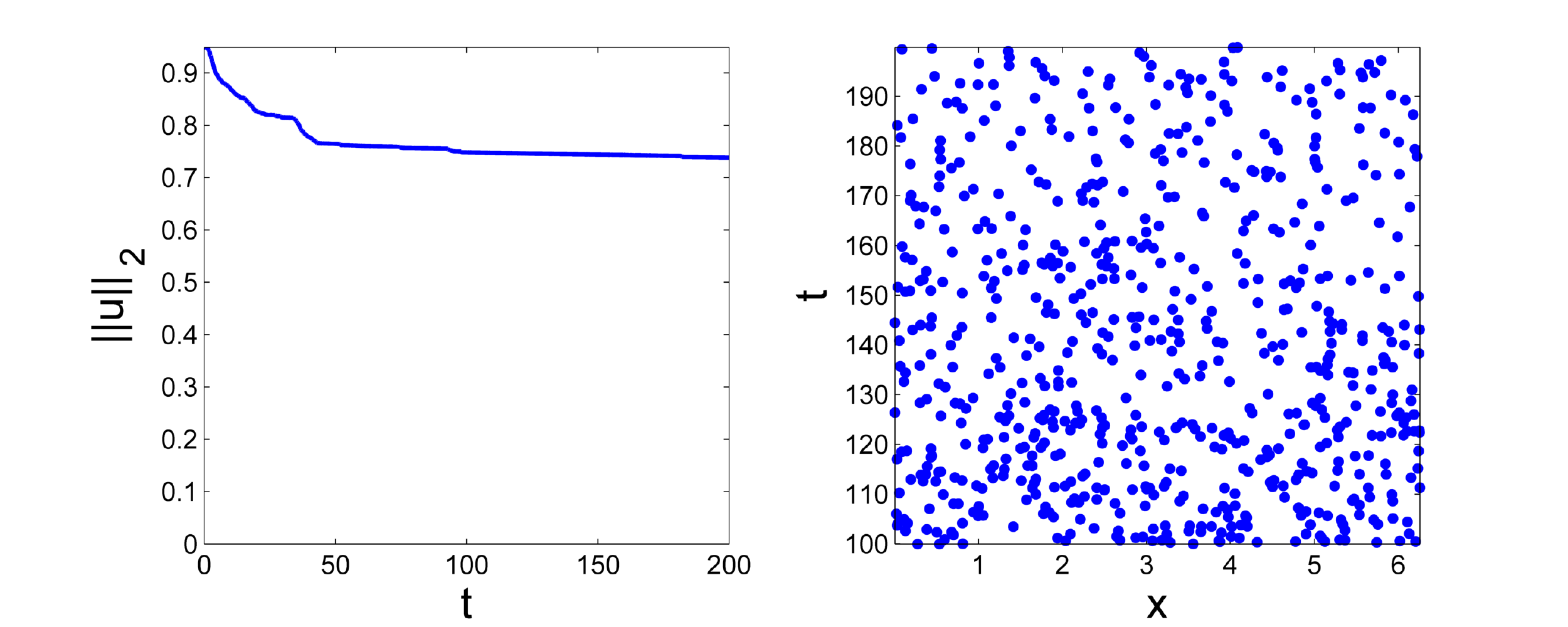}
\caption{Left: Decay of $L^{2}$ norm of the solution: after 100 time units
the decay rate becomes small. Right: Locations of extreme events in
an ensemble of simulations.}

\label{fig:energyDecay} 
\end{figure}

For the statistical studies performed in this work, we evolve a sum
of 31 complex exponentials with independent, uniformly distributed
random phases. We compute statistics by averaging over time and space
over 300 ensembles, each spanning 100 time units ($t=100$ to $t=200$).
There is no external forcing in our simulations and all the energy
of the system comes through the initial conditions while dissipation
occurs whenever an extreme event takes place. To this end, we do not
observe an exact statistical steady state in our simulations, but after
an initial transient where a moderate amount of energy is dissipated
through selective damping, the solution settles to a nearly (or very
slowly varying) statistical steady state where the $L^{2}$ norm decays
slowly (see Figure~\ref{fig:energyDecay}). We focus on this slowly
varying regime where roughly 2-4 extreme events occur per simulation
in the time window $t\in[100,200]$, and, as shown by Figure~\ref{fig:energyDecay},
these extreme events are uniformly prevalent (roughly) throughout
this time window (although they are slightly more common for earlier
times).

\section{Nonlinear instabilities induced by spatially localized energy\label{sec:energyLocalization}}

In this section we examine the role of \textit{spatially} \emph{localized}
energy in the formation of an extreme event. More specifically, we
define the energy $E$ of a solution as 
\[
E\triangleq r^{2}=\int\left|u^{2}\left(x\right)\right|dx,
\]
where $r$ is the $L^{2}$ norm of the solution, which is conserved
by undamped MMT \citep{zakharov2001}. In the undamped MMT equation,
localized initial data with energy above some critical level leads
to a finite time blowup \citep{cai2001,zakharov2001}. Here we examine
how this critical energy level varies with the degree of initial energy
localization, as well as the energy of the background state, in the
presence of selective dissipation. Both of these parameters are important
to determine the critical scale that is most sensitive for the formation
of an extreme event. For the zero background case, we are able to
analytically determine this relationship by deriving a scale invariant
family of solutions. We investigate the non-zero background case numerically.

\textbf{Zero background energy: scale invariant solutions.} We begin
our analysis by focusing on localized perturbations when we have zero
energy background in the system. More specifically, we consider a
family of initial data of the form $u(x,0)=u_{0}(x;c,L)=ce^{-2(x/L)^{2}}$
and determine how the critical energy level required for blow-up depends
on the length scale $L$. To do so, we derive an $L$-parametric family
of solutions $w_{L}$, $L>0$, defined by the scaling of a given solution
$u(x,t)$ 
\begin{align*}
w_{L}(x,t)=\frac{1}{L^{p}}u\left(\frac{x}{L},\frac{t}{L^{q}}\right).
\end{align*}
To determine $p$ and $q$, we plug this anzatz into MMT with no dissipation,
which gives: 
\begin{align*}
\frac{i}{L^{p+q}}u_{t}=\frac{1}{L^{p+\alpha}}|\partial_{x}|^{\alpha}u+\frac{\lambda}{L^{3p-\beta}}|\partial_{x}|^{-\beta/4}\left(\left\vert |\partial_{x}|^{-\beta/4}u\right\vert ^{2}|\partial_{x}|^{-\beta/4}u\right).
\end{align*}
So $w$ is also a solution to MMT if $q=\alpha$ and $p=(\alpha+\beta)/2$,
giving us the following family of solutions: 
\begin{align*}
w_{L}(x,t)=\frac{1}{L^{(\alpha+\beta)/2}}u\left(\frac{x}{L},\frac{t}{L^{\alpha}}\right)
\end{align*}
Therefore, if for a reference lengthscale $L=1,$ we have the critical
energy norm $r_{crit}\left(1\right)$ (associated with an initial
condition $u_{0}\left(x;c^{*},1\right)$) that leads to a blow-up
solution, the corresponding critical energy for initial data localized
for an arbitrary length scale $L$ will be 
\[
r_{crit}^{2}(L)=\frac{1}{L^{\alpha+\beta}}\int u_{0}^{2}\left(\frac{x}{L};c^{*},1\right)dx=L^{1-\alpha-\beta}r_{crit}^{2}(1).
\]
Hence, the critical energy norm $r_{crit}\left(L\right)$ required
to initiate a blow-up is given by 
\begin{align}
r_{crit}(L)=L^{(1-\alpha-\beta)/2}r_{crit}(1).\label{eq:genCritE}
\end{align}
We consider the special case $\alpha=1/2,\beta=0$, which gives 
\begin{equation}
r_{crit}(L)=\sqrt[4]{L}r_{crit}(1).\label{critical_energy_undamped}
\end{equation}
Since the above function decreases to 0 as $L$ becomes small, in
the deep water wave dispersion case only a \emph{small amount of localized
energy} is sufficient to initiate a blow-up. This fact holds as long
as the exponent of $L$ in (\ref{eq:genCritE}) is positive, meaning
$\beta<1-\alpha$, or simply $\beta<1/2$ using the standard value
of $\alpha=1/2$. Pushkarev and Zakharov \citep{pushkarev2013} use
$\beta=-3$ to study extreme waves, which is small enough to ensure
that this relationship between localization and energy criticality
still holds. In fact, with $\beta=-3$ we have $r_{crit}(L)=L^{7/2}r_{crit}(1)$,
so this relationship ($r_{crit}$ decreasing as $L$ decreases) would
presumably be even stronger than the $\beta=0$ case we consider.

Note that for the case that selective dissipation is present, very
localized amounts of energy will be rapidly dissipated. In particular,
if energy is \emph{too} localized, then the selective Laplacian damping
is dominant compared with the instability of the nonlinear terms and
the amplitude of $u$ decreases relative to its initial state. However,
for values of $L$ that are not excessively small, we have a rapid
growth of the amplitude of $u$ that leads to an energy cascade (see
next section) to smaller scales and subsequent dissipation by the
selective Laplacian. In this way the high wavenumber damping prevents
the formation of a singularity due to continuous energy transfer and
accumulation to infinitesimally small scales and results in a finite
lifetime for the extreme event (Figure \ref{fig:exampleEE}).

\begin{figure}[ptb]
\centering \includegraphics[width=4.8533in,height=2.9542in]{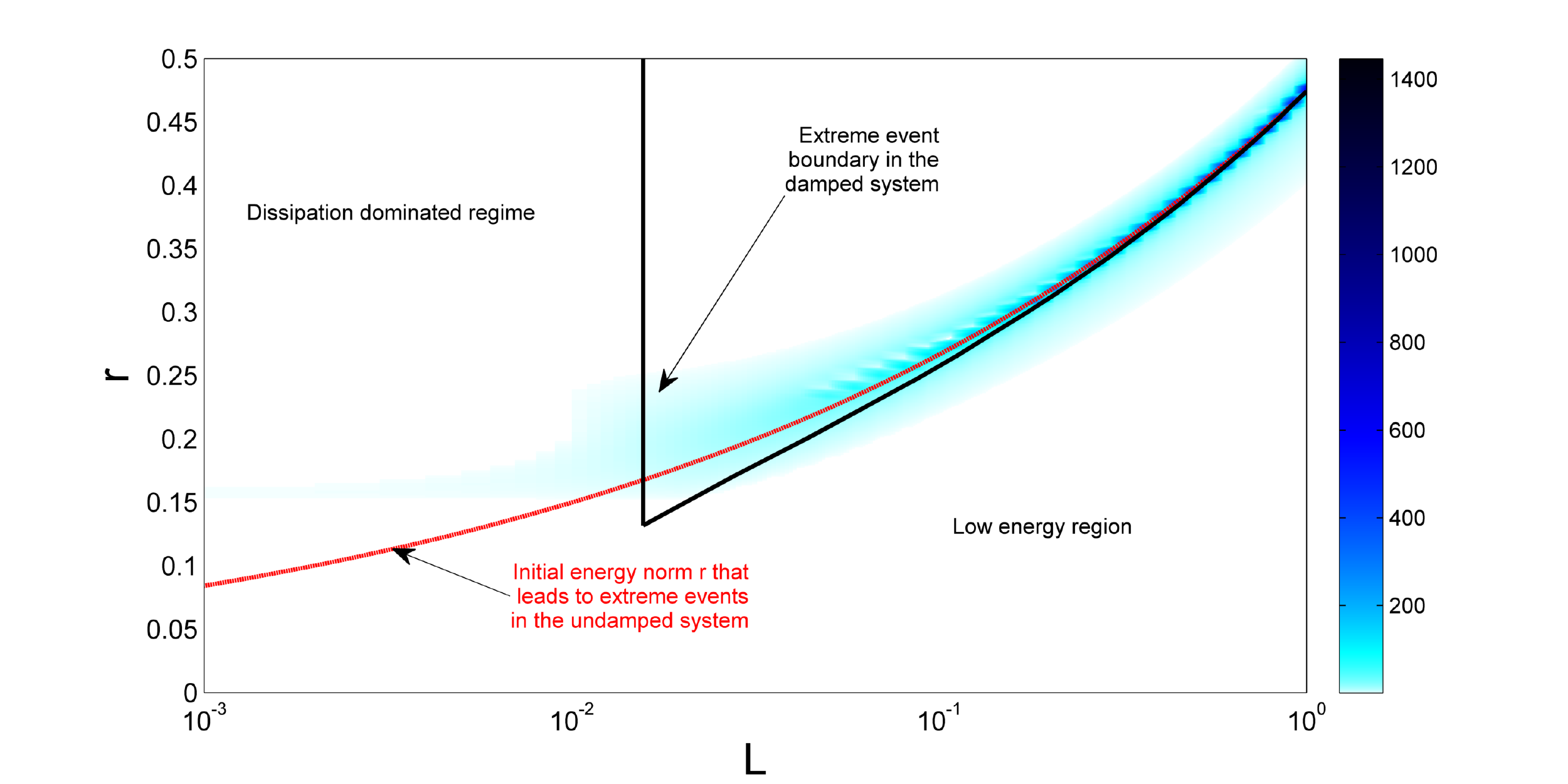}\caption{Critical energy norm of a localized perturbation that leads to the
formation of an extreme event for the undamped (red dashed curve)
and the damped MMT model in the absence of background energy. The
latter is described in terms of the finite-time divergence $q$ of
nearby trajectories (color map) and the maximum value of the response
field $\left\vert u\right\vert .$}

\label{fig:MMTCritEnergy} 
\end{figure}

Therefore, in the damped MMT, for each localization scale $L$ that
is not excessively small we expect there to be a critical amount of
energy that will trigger a nonlinear instability resulting in an extreme
event. We expect that, except for excessively small values of $L$,
the above analysis will still hold and the dissipation will only become
relevant in the late stages of an extreme event where it prevents
the formation of a singularity. We quantify the critical energy for
the damped system using two different measures. First, we compute
the finite-time divergence of nearby (in terms of energy) initial
perturbations through the quantity 
\begin{align*}
|\partial_{r}q(r,L)|\triangleq\left\vert \frac{\partial}{\partial r}\frac{\max_{x,t}|u(x,t;r)|}{\max_{x}u(x,0;r)}\right\vert 
\end{align*}
This quantity is displayed by a color plot in Figure~\ref{fig:MMTCritEnergy}.
We use the sharp ridge of $|\partial_{r}q|$ to determine the critical
energy level at which the transition to extreme events occurs. Additionally,
we determine the critical energy level by determining the set of values
$(r,L)$ at which $q(r,L)>1.5$. The black curve in Figure~\ref{fig:MMTCritEnergy}
outlines the region where $q$ exceeds this threshold value. This
curve compares favorably with the results from the first method. Also
in Figure~\ref{fig:MMTCritEnergy} we present with a red dashed curve
the critical energy norm for the undamped system, given by (\ref{critical_energy_undamped}).
We emphasize that even though Figure~\ref{fig:MMTCritEnergy} was
generated by numerically solving (\ref{eq:MMT}) on a domain of size
$16\pi$ with periodic boundary conditions, these results do not
change if the domain size is increased further due to the localization
of these examples. This behavior contrasts sharply with similar experiments
of the nonlinear Schrodinger equation, where the values of $|q|$
never become large and the sharp gradient seen in Figure~\ref{fig:MMTCritEnergy}
does not occur.

We note that for the case of the damped system the critical energy
norm closely resembles the analytical prediction (\ref{critical_energy_undamped}),
which is a result of the interplay between dispersion and nonlinearity.
This is the case until we reach the critical scale $L_{c}$, below
which dissipation is important and no extreme solutions can occur.
To this end this spatial scale $L_{c}$ is the most sensitive to localized
perturbations i.e. it can be triggered with the lowest amount of energy,
and it is essentially the smallest scale where dissipation is still
negligible. The existence of this critical scale that triggers extreme
events is the result of the synergistic action of dispersion, nonlinearity,
and small scale dissipation. 

\textbf{Case of finite background energy. }We now consider the formation
of an extreme event out of a background with non-zero energy, that
is, we evolve initial data of the form $u(x,0)=b+ce^{-2(x/L)^{2}}$.
We first consider the case of small ratio $\frac{c}{b}\ll1$ where
we can investigate the evolution of $u$ by performing a linearized
stability analysis about the plane wave solution of MMT, $u(x,t)=be^{-i\lambda b^{2}t}$.

For the nonlinear Schrödinger equation with periodic boundary conditions,
this plane wave solution is unstable to Fourier mode perturbations
of wavenumber $n$ when the following condition is satisfied: 
\[
n<\frac{L_{x}}{2\pi}\sqrt{-2\lambda b^{2}},
\]
where $L_{x}$ is the domain width. The above is known as the Benjamin-Feir
instability and has been studied extensively by many authors \citep{benjaminFeir1967,benjamin1967,zakharov1968,yuen1978}.

In the context of the undamped MMT equation, the Benjamin-Feir
instability can be generalized. In particular one can show that for
the case $\beta=0$, the plane wave solution is unstable if $\lambda<0$
and 
\begin{equation}
n<\frac{L_{x}}{2\pi}\left(-2\lambda b^{2}\right)^{1/\alpha}.\label{eq:genBF}
\end{equation}
Clearly, for $\alpha=2$, the above result agrees exactly with the
classical Benjamin-Feir instability criterion of NLS.

\begin{figure}[ptb]
\centering \includegraphics[width=4.7305in,height=3.3217in]{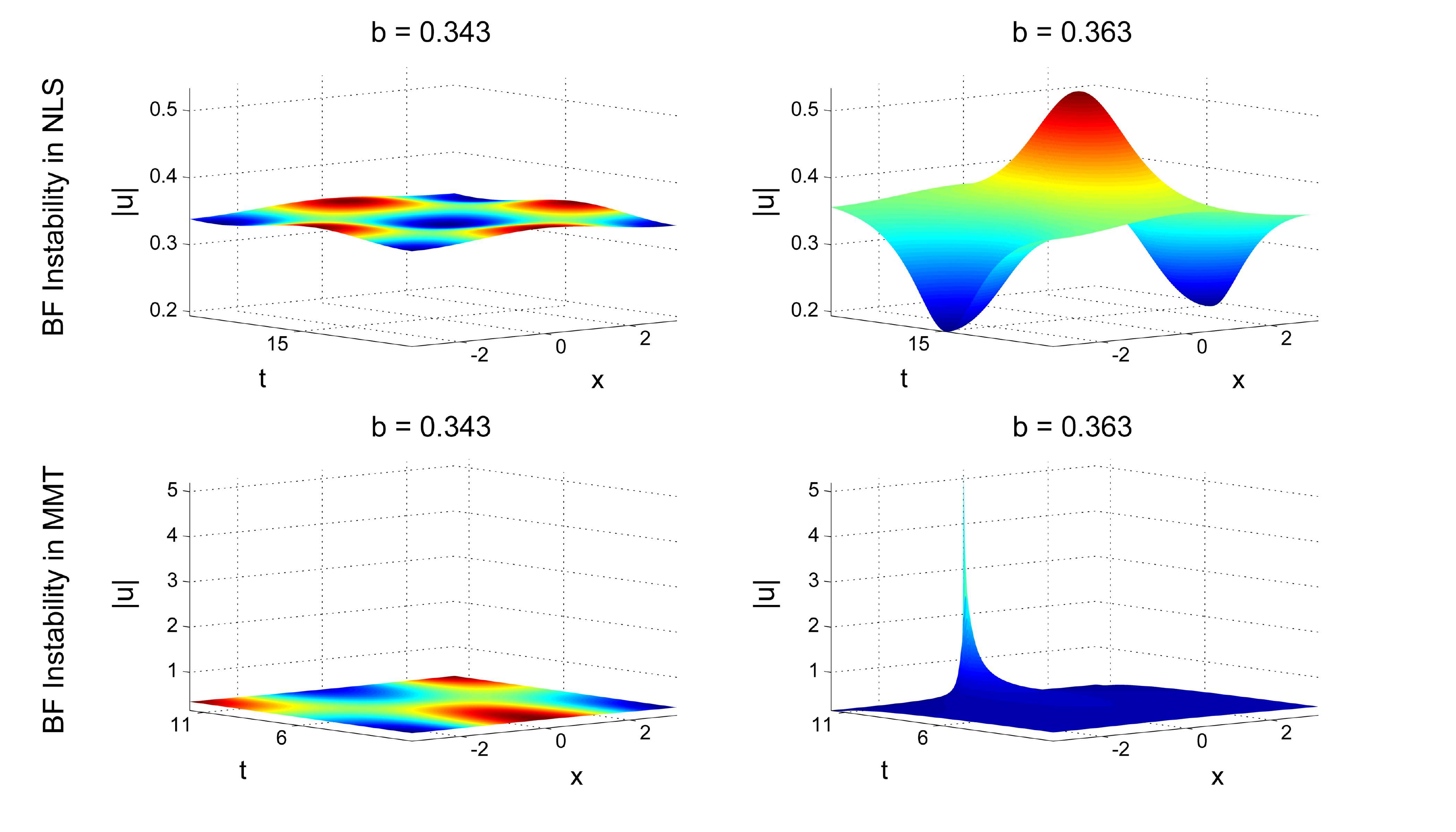}\caption{Benjamin-Feir instability for the NLS equation (top) and the MMT model
(bottom) with $\left(\alpha=0.5,\beta=0,\lambda=-4\right).$}

\label{fig:benjaminFeir} 
\end{figure}

We emphasize that although the Benjamin-Feir modulation instability
is present in both the focusing MMT and the NLS, its manifestation
is not the same in each case. We illustrate this fact by numerical
experiments involving no selective damping. For both equations, we
take $\lambda=-4$ and $L_{x}=2\pi$, meaning that the critical value
of $b$ in (\ref{eq:genBF}) is roughly 0.35. Values of $b$ larger
than this admit at least one unstable mode, and positive $b$ less
than this value have no unstable modes. We evolve initial data of
the form $u(x,0)=b+\epsilon\cos(x)$ with $b\approx0.34$ and $b\approx0.36$.
For each value of $b$, we set $\epsilon=0.01$. When $b\approx0.34$,
the small perturbation does not grow for the MMT or the NLS, agreeing
with the linearized analysis (see Figure~\ref{fig:benjaminFeir}).
For the NLS, unstable perturbations of this kind initiate a nearly-periodic
orbit where large, but bounded, coherent structures repeatedly appear
and subsequently dissolve in a Fermi-Pasta-Ulam-like recurrent cycle
\citep{ablowitz1990,yuen1978}. However, in the MMT, these unstable
perturbations grow continuously and collapse into a singularity in
finite time (see Figure~\ref{fig:benjaminFeir}). This mechanism
of collapse initiation via modulation instability, which has also
been studied by Cai et. al. \citep{cai2001}, has significant implications
for our critical energy analysis for the nonzero background case.
Although including high frequency damping will prevent the formation
of a singularity, such damping will not prevent an extreme event from
occurring since it only becomes relevant after energy has been transferred
to the small scales; that is, after an extreme event has already occurred
(as in Figure~\ref{fig:exampleEE}). Thus, if $b$ and $L_{x}$ satisfy
the Benjamin-Feir instability criterion (\ref{eq:genBF}) with $n=1$,
initial conditions of the form $u(x,0)=b+ce^{-2(x/L)^{2}}$ will initiate
an extreme event for any $c>0$.

\begin{figure}[ptb]
\centering \includegraphics[width=5.7486in,height=3.3425in]{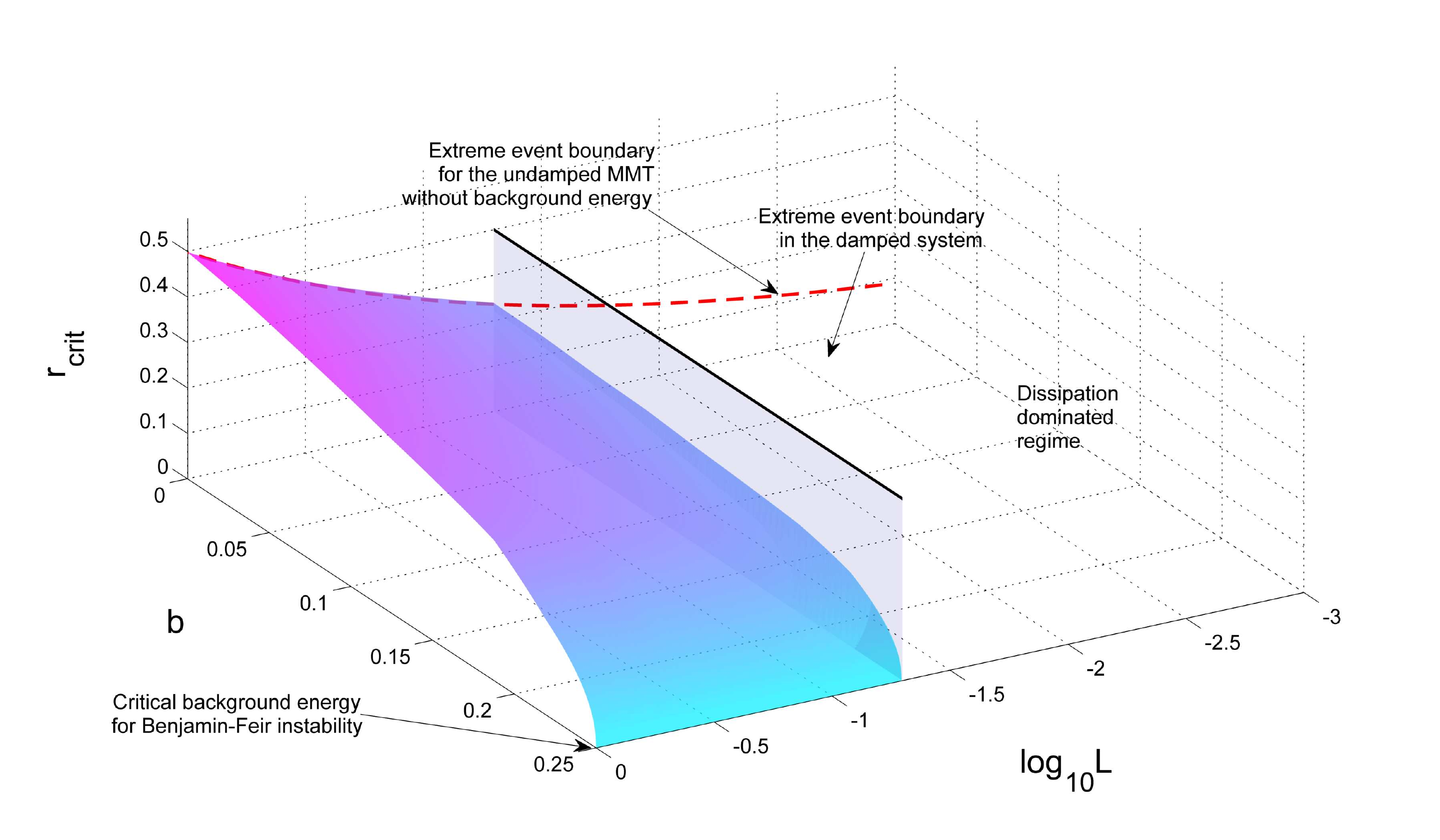}\caption{Critical energy norm of localized perturbations that lead to extreme
events in the presence of background energy for the damped MMT model
$\left(\alpha=0.5,\beta=0\right)$.}

\label{critical_surf} 
\end{figure}

Figure~\ref{critical_surf} displays the critical energy norm of
the perturbation $ce^{-2(x/L)^{2}}$ required to initiate an extreme
event for various values of $b$ and $L$. For $b=0$, this critical
amplitude is precisely the curve described above and displayed in
Figure~\ref{fig:MMTCritEnergy}, and for large enough $b$ this critical
amplitude is infinitesimal due to the Benjamin-Feir instability. For
intermediate values of $b$, the critical amplitude presents a smooth
transition between the two theoretically understood regimes. For $b=0$,
we noted previously that the more localized the energy is, the smaller
amount of this energy is required to initiate a blowup. This fact
remains true for nonzero background energy $b$ until the point where
$b$ is large enough so that a Benjamin-Feir instability occurs, which
in this case ($L_{x}=8\pi,\lambda=-4,\alpha=1/2$) occurs at $b=0.25$.
picture is another manifestation that we can have extreme responses
well below the Benjamin-Feir energy threshold as a result of the interplay
between nonlinearity, dispersion and dissipation. 

\section{Stochastic dynamics during an extreme event\label{sec:gaborStats}}

So far we have examined the conditions that lead to extreme wave solutions.
In this section we will study the nonlinear interactions taking place
during the occurrence of an extreme event using tools from stochastic
analysis. We choose to project the solution $u$ onto an appropriate
(localized) set of modes. Given the localized character of extreme
events, global basis elements such as Fourier modes will not be able
to describe effectively their dynamical properties since, despite
their large amplitude, extreme events carry very small portion of
energy of the overall field spectrum.

To this end, it is more informative to choose a set of modes which
incorporate the localized character of the extreme events. We use
the following family of Gabor basis elements consisting of complex
exponentials multiplied by Gaussian window functions: 
\begin{equation}
v_{n}(x;x_{c})\triangleq\exp\left[-2\frac{d(x,x_{c})^{2}}{L^{2}}\right]e^{i2\pi nx/L},\text{ \ \ }n=0,1,2,...\label{eq:gaborBasis}
\end{equation}
where $d(x,x_{c})=\min(|x-x_{c}|,2\pi-|x-x_{c}|)$ expresses the distance
from the center point $x_{c}$ measured in the periodic domain. We
then compute the Gabor projection coefficients of the solution. 
\[
Y_{n}(x_{c},t)\triangleq\langle u(x,t),v_{n}(x;x_{c})\rangle/||v_{n}(x;x_{c})||^{2},
\]
where $\langle\cdot,\cdot\rangle$ denotes the standard $L^{2}$ inner
product. We noted in Section~\ref{sec:energyLocalization} that there
is a critical scale $L_{c}\approx0.01$ that is most sensitive to
the formation of an extreme event, in that the required energy to
trigger an extreme event is smallest at this particular scale. In
practice, we have observed that these extreme events typically originate
by energy localization in a slightly larger scale than $L_{c}$. This
motivates our choice of $L=\pi/100\approx3L_{c}$, which is still
extremely sensitive to small perturbations (Figure~\ref{fig:MMTCritEnergy}).
After a sufficient amount of energy is localized in this scale, an
extreme peak is then produced as energy is transferred into the smaller
scales until it reaches the scale at which selective dissipation is
present. We have chosen $L=0.031$ based on this argument with the
goal of using the coefficient $Y_{0}$ as an indicator of an upcoming
extreme event.

\begin{figure}[ptb]
\centering \includegraphics[width=5.2641in,height=3.0329in]{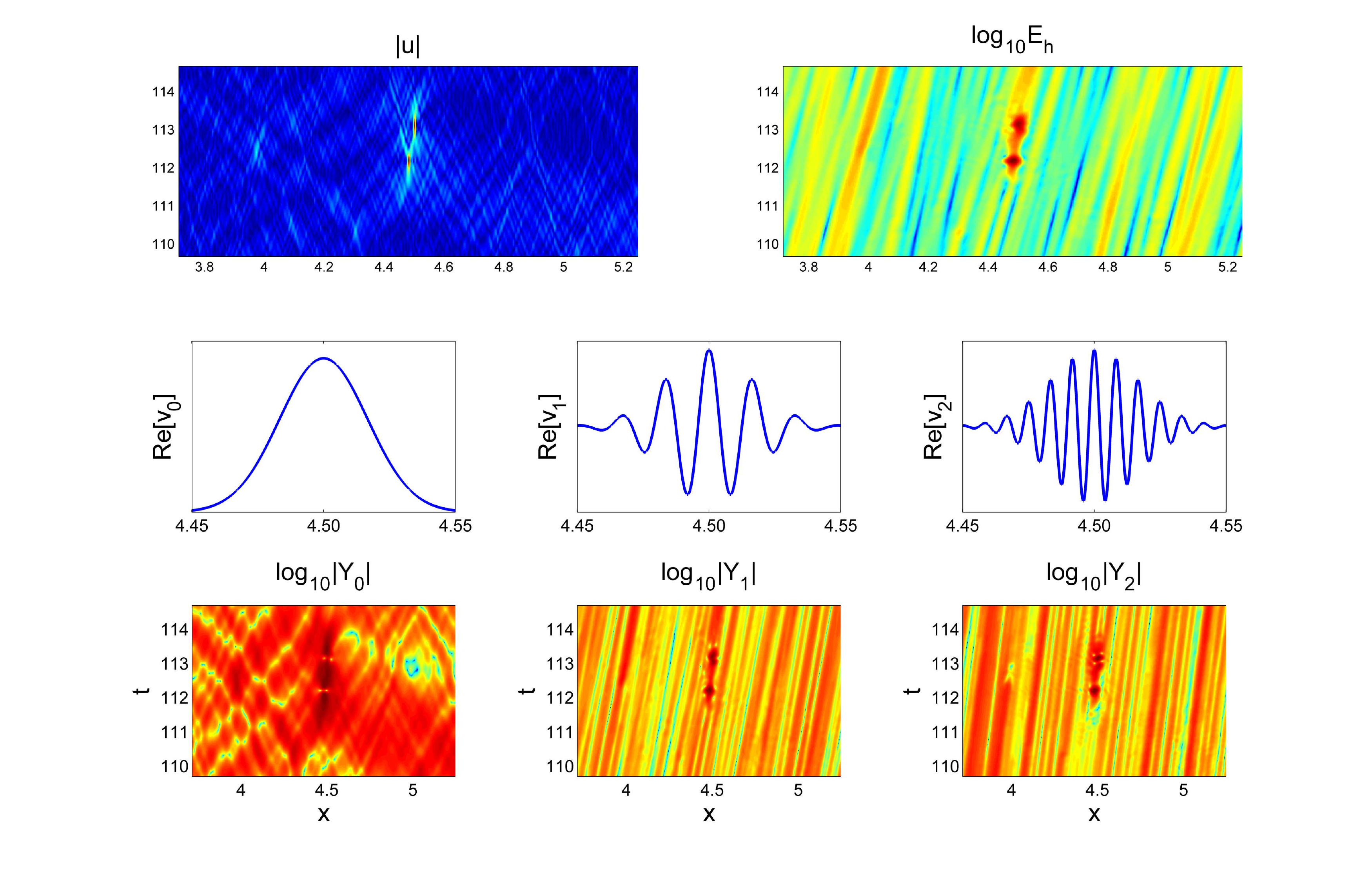}\caption{Energy cascade during an extreme event. Top row: extreme event shown
in $\left(x,t\right)$ space together with the high wavenumber energy
$E_{h}=\sum_{n\geq1}\left\Vert Y_{n}\right\Vert ^{2}.$ Second row:
Gabor basis elements $v_{n}(x;0).$ Third row: modulus of the Gabor
coefficients $\left\Vert Y_{n}(x_{c},t)\right\Vert .$}

\label{fig:localBasis} 
\end{figure}

In Figure \ref{fig:localBasis} we present the Gabor coefficients
in space and time during the occurrence of a pair of extreme events.
More specifically, in the first row (left) we show the extreme waves
in the $\left(x,t\right)$ space. The Gabor basis elements are shown
in the second row and the Gabor coefficients are shown in the third
row. The Euclidian sum of the oscillatory Gabor coefficients, $E_{h}=\sum_{n\geq1}\left\Vert Y_{n}\right\Vert ^{2}$
expresses the energy that `lives' in high wavenumbers and this is
shown in the top-right panel.

For the Gabor coefficients that correspond to the oscillatory basis
elements ($Y_{1},Y_{2}$), we note that away from the region of the
extreme events, wave components propagate almost independently, according
to the dispersion relation, in a close to linear fashion. The energy
of the non-oscillatory mode (expressed through the Gabor coefficient
$Y_{0}$) presents a more static (non-propagating) behavior with high
intensity that builds up before the extreme event. During the strongly
non-linear phase (of the extreme response), the Gabor coefficients
of the oscillatory basis elements are not governed by the dispersion
relation anymore, but they also present a more static (non-propagating)
behavior characterized by a large build up of their energy. This is
the result of a strong, nonlinear energy cascade initiated from the
unstable lengthscale $L$, described by the non-oscillatory mode $v_{0},$
and ending to higher wavenumbers where it is dissipated (Figure \ref{fig:localBasis}).

\begin{figure}[ptb]
\centering \includegraphics[width=5.2926in,height=2.1447in]{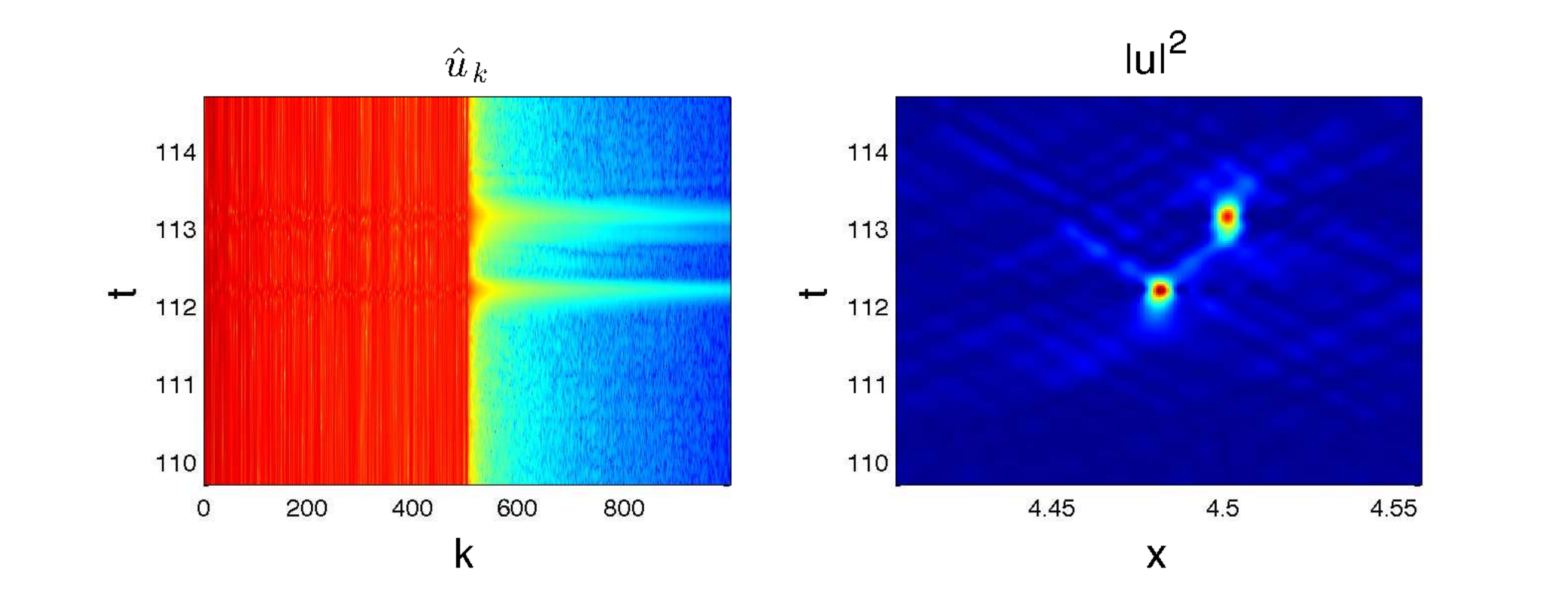}\caption{Energy cascade to higher wavenumbers during the occurrence of extreme
events shown in terms of the moduli of the Fourier coefficients for
the MMT model.}

\label{fig:fourierCoefs} 
\end{figure}

\textbf{Statistics and energy transfers during the dissipation phase.}
The strong nonlinear energy transfer from the unstable scale $L$
to smaller scales is manifested by the non-Gaussian statistics of
the Gabor coefficients during an extreme event. The connection between
nonlinear energy transfers and non-Gaussian statistics has been rigorously
established in \citep{sapsis11a,sapsis_majda_mqg,sapsis_majda_tur}
in the context of viscous turbulent flows. Consider an orthonormal set of modes $v_{i},$
$i=0,1,2,...,n_{c}$ (although the modes (\ref{eq:gaborBasis}) are
not orthonormal, they are nearly so) which are active during the occurrence
of an extreme event and and let the background stage (i.e. the full
stochastic solution during a non-extreme regime) described by $u\left(x,t\right).$
Then, the MMT equation with $\alpha=0.5,\beta=0$ 
\[
u_{t}=-i\left\vert \partial_{x}\right\vert ^{1/2}u-i\lambda\left\vert u\right\vert ^{2}u+Du
\]
will take the projected form for each $v_{i}$ 
\begin{align}
\frac{dY_{i}}{dt} & =\left\langle \left[D-i\left\vert \partial_{x}\right\vert ^{1/2}\right]u,v_{i}\right\rangle +\sum_{k}Y_{k}\left\langle \left[D+\left\vert \partial_{x}\right\vert ^{1/2}\right]v_{k},v_{i}\right\rangle \label{Energy_eq}\\
 & -i\lambda\left\langle \left\vert u+\sum_{k}Y_{k}v_{k}\right\vert ^{2}\left(u+\sum_{k}Y_{k}v_{k}\right),v_{i}\right\rangle .\nonumber 
\end{align}
The growth rate of the energy of $Y_{i}$ will have the form 
\[
\frac{d\left\vert Y_{i}\right\vert ^{2}}{dt}=\frac{dY_{i}}{dt}Y_{i}^{\ast}+\frac{dY_{i}^{\ast}}{dt}Y_{i}=2\operatorname{Re}\left[\frac{dY_{i}}{dt}Y_{i}^{\ast}\right].
\]
Note that in equation (\ref{Energy_eq}) the first line on the right hand side 
involves either energy conserving terms such as wave dispersion or
negative definite terms such as dissipation (which occurs for high
wavenumbers only). All the other contributions towards changes of
$\left\vert Y_{i}\right\vert ^{2}$ will only occur through the nonlinear
interactions of the modes: 
\[
\left.\frac{d\overline{\left\vert Y_{i}\right\vert ^{2}}}{dt}\right\vert _{NL}=-2\operatorname{Re}\left[\overline{i\lambda\left\langle \left\vert u+\sum_{k}Y_{k}v_{k}\right\vert ^{2}\left(u+\sum_{k}Y_{k}v_{k}\right),v_{i}\right\rangle Y_{i}^{\ast}}\right]
\]
where the bar denotes ensemble average. We focus on the energy cascade
regime from the mode $v_{0}$ to higher wavenumber modes during an
extreme event.

\begin{figure}[ptb]
\centering \includegraphics[width=5.572in,height=2.3047in]{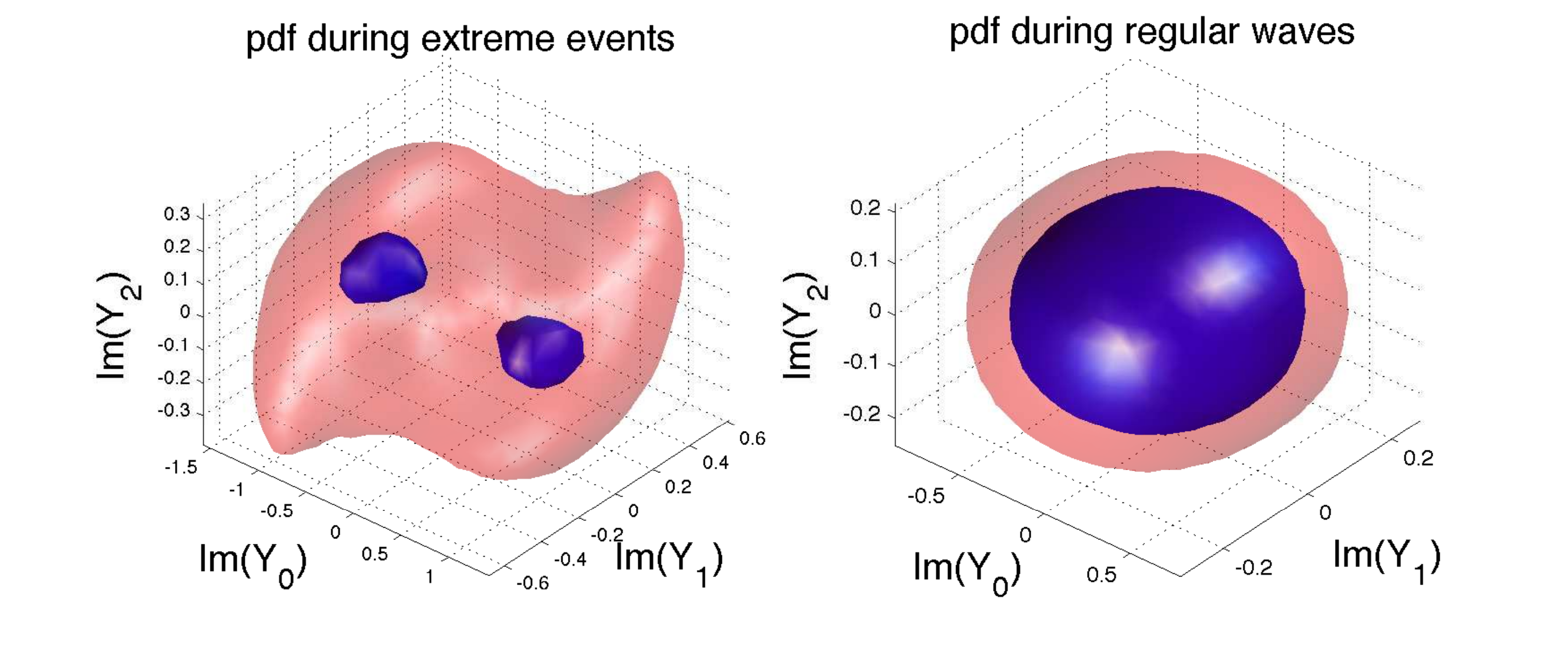}\caption{Isosurfaces of the joint pdf of the imaginary part of $Y_{0},Y_{1},$
and $Y_{2}$ near to (left) and away from (right) extreme events.}

\label{fig:gaborJointPDF} 
\end{figure}

We compute the Gabor coefficients for 400 values of $x_{c}$ equally
distributed between 0 and $2\pi$, with $L\approx3L_{c}.$. We then
classify each point $(x_{c},t)$ into two regimes: points nearby an
extreme event and points away from extreme events. For each of these
groups, we compute the joint statistics of the Gabor coefficients
using data from an ensemble of MMT simulations with random initial
data (details of these simulations are given in Section~\ref{sec:simulation_statistics}).
Figure~\ref{fig:gaborJointPDF} displays the joint statistics of
the imaginary parts of $Y_{0},Y_{1}$, and $Y_{2}$ near (left subplot)
and far (right subplot) from extreme events.

Away from extreme events the isosurfaces of the probability density
function are elliptical, indicating that the Gabor coefficients are
nearly Gaussian in this regime. Moreover, the time oscillatory character
of the wave components results in zero average value (in the ensemble
sense) for all the corresponding Gabor coefficients $\overline{Y_{i}}=0,$
$i=1,2,...$ (for both regimes). Due to this fact, as well as the
Gaussian distribution of the coefficients $Y_{i}$ in the non-extreme
events regime, the average change of their energy due to nonlinear
interactions, becomes zero.

On the other hand, the statistics near extreme events are highly non-Gaussian,
with $Y_{0}$ exhibiting a bimodal distribution. The real parts of
the Gabor coefficients are distributed similarly. This non-Gaussian
distribution is directly related to the energy cascade from the non-oscillatory
mode to the strongly dissipative, high wavenumber modes. 

\begin{figure}[ptb]
\centering \includegraphics[width=3.5734in,height=2.4639in]{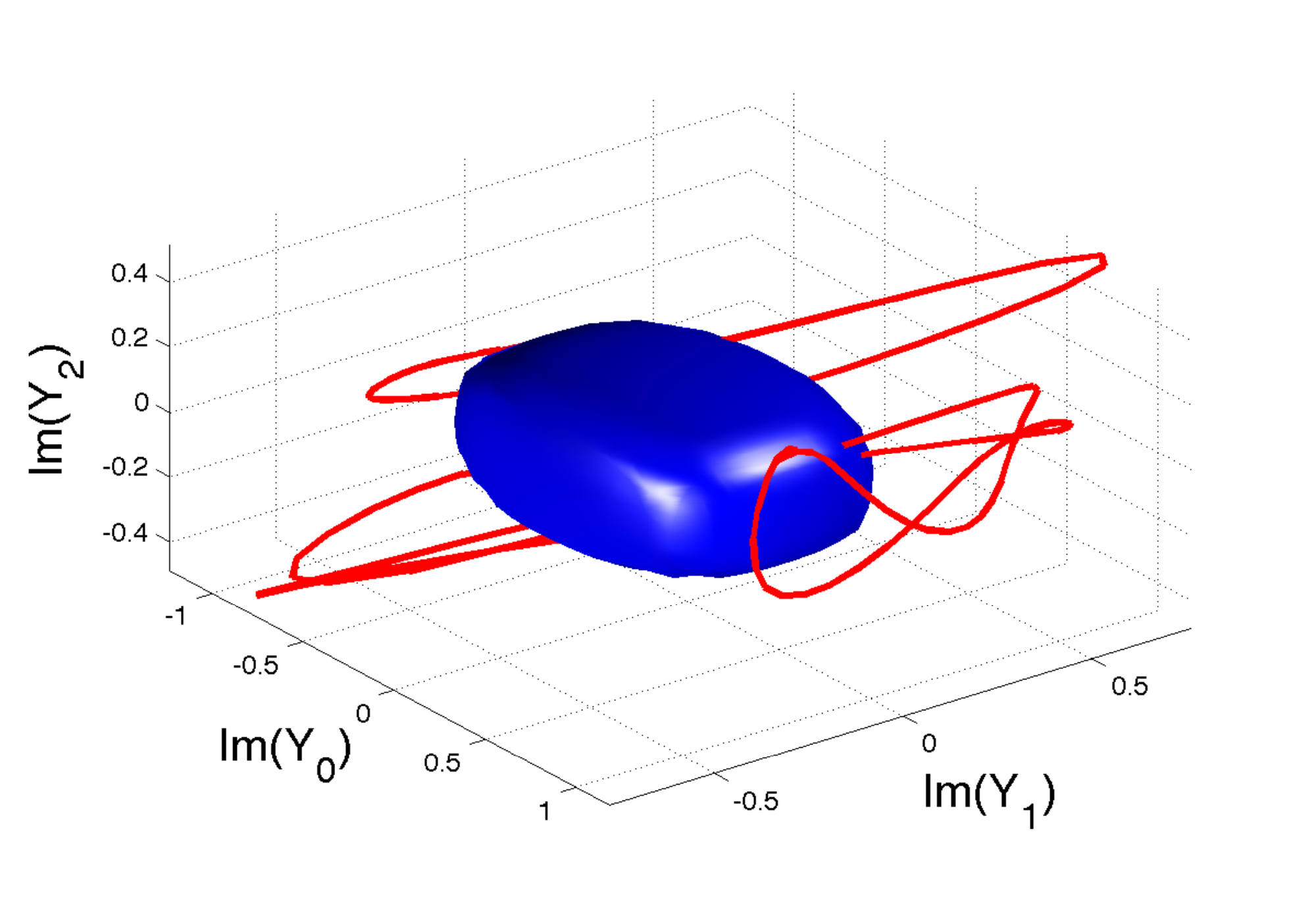}\caption{The red curves show trajectories of the imaginary parts of $Y_{0}$,
$Y_{1}$, and $Y_{2}$ during extreme events. The blue surface is
an isosurface of the joint density function for the imaginary parts
of $Y_{0}$, $Y_{1}$, and $Y_{2}$ containing 97\% of the total probability--this
shape is dominated by the Gaussian random waves that propagate with
random phase under the effect of dispersion and weak nonlinearity.}

\label{extreme_event_gaussian_back} 
\end{figure}

\begin{figure}[ptb]
\centering \includegraphics[width=4.932in,height=2.5624in]{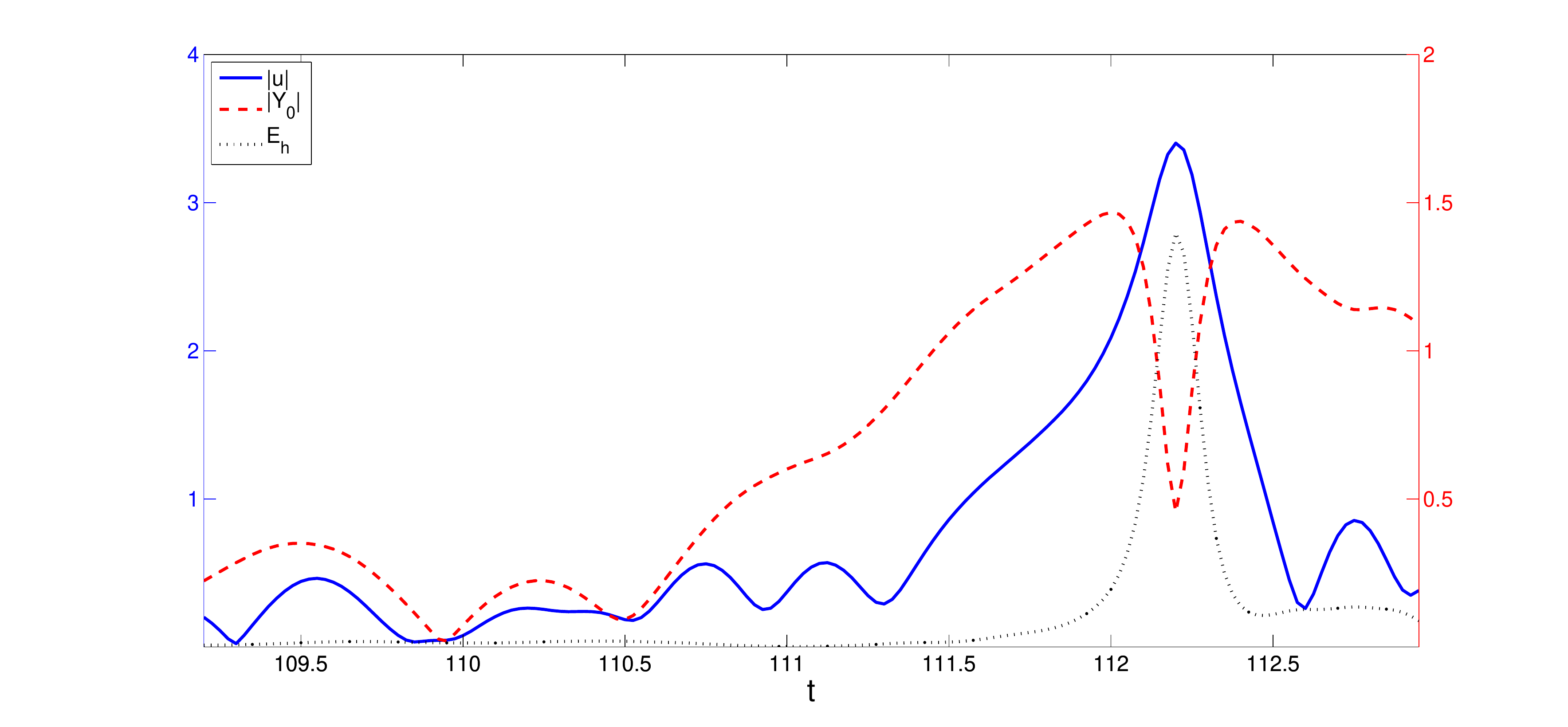}\caption{Time evolution of $|u|$, $|Y_{0}|$, and $E_{h}$ at a spatial location
of an extreme event. The y-axis scale on the left corresponds to $|u|$,
the right y-axis scale corresponds to $|Y_{0}|$ and $E_{h}$.}

\label{fig:gaborTimeSeries} 
\end{figure}

\textbf{Dynamics during the built-up phase of the extreme event.}
In Figure \ref{extreme_event_gaussian_back}, we show how an extreme
event trajectory emerges out of the Gaussian background describing
the heat bath of waves propagating under the dominant effect of the
dispersion relation. From the same figure it is clearly illustrated
how extreme events are associated with large values of $|Y_{0}|$.
The nature of this association is particularly interesting: $|Y_{0}|$
becomes large just \emph{before} (and after) extreme events. An example
of this behavior is displayed in Figure~\ref{fig:gaborTimeSeries},
where we observe the increase of $|Y_{0}|$ while the overall response
field $\left\vert u\right\vert $ has regular values. This growth
continues until we have an extreme event and it is followed by a sudden
drop that is associated with an energy transfer to high wavenumbers
illustrated by the energy $E_{h}$ (as described previously). This
agrees with observations by Cai et. al. \citep{cai2001} that these
extreme events form by focusing energy to high wavenumbers until saturation
at a critical scale, at which point they radiate energy back to larger
scales. We are particularly interested in the predictive utility of
the localized energy $\left|Y_{0}\right|$ buildup that occurs before
extreme events, often 1-2 time units in advance for the considered
set of parameters. 

This phenomenon agrees with our analysis from Section~\ref{sec:energyLocalization},
where we showed that a sufficient amount of localized energy is sufficient
to trigger an extreme event. These localizations of energy occur randomly
through the dispersive propagation of waves that have random phases.
Due to the localized nature of $v_{0}$, the associated Gabor coefficient
$|Y_{0}|$ measures such localized energy and is thus an indicator
of an extreme event in the near future. When the extreme event occurs,
energy is transferred into the more oscillatory modes ($v_{1},v_{2},\dots$),
but the Gabor coefficients associated with these modes lack predictive
utility since they grow simultaneously with, rather than prior to,
the extreme event (see Figure~\ref{fig:gaborTimeSeries}). 

\section{Short-term prediction of extreme events\label{sec:prediction}}

The Gabor coefficient $Y_{0}$ is a measure of energy localized at
a particularly sensitive length scale, at which only a small amount
of energy is necessary to trigger an extreme event. Thus, large values
of $|Y_{0}|$ often indicate that an extreme event will occur in the
near future. We now use this fact to develop short-term predictive
capacity for extreme events. To do so, we first compute, for various
values of $\mathcal{Y}_{0}$, the following family of probability distributions:
\begin{align}
F_{\mathcal{Y}_{0}}(\mathcal{U})\triangleq\mathcal{P}\left(\max_{\substack{|x^{*}-x_{c}|<L\\
t^{\ast}\in\lbrack t-1.5,t+1.5]
}
}|u(x^{\ast},t^{\ast})|>\mathcal{U}\text{ \ }\Big|\text{ }|Y_{0}(x_{c},t)|=\mathcal{Y}_{0}\right).\label{eq:Y0CondPDF}
\end{align}
That is, given a particular value of $\left|Y_{0}(x_{c},t)\right|$,
we compute the probability that $|u|$ exceeds $\mathcal{U}$ nearby.
The timescale of 1.5 time units has been chosen based on our observation
of the time required for the transition from large values of $\left\vert Y_{0}\right\vert $
to extreme values of $\left\vert u\right\vert $ \ We compute these
distributions from an ensemble of 300 simulations, each spanning 100
time units (see Section~\ref{sec:simulation_statistics} for simulation
details). 
\begin{figure}[ptb]
\centering \includegraphics[width=5.7406in,height=2.1759in]{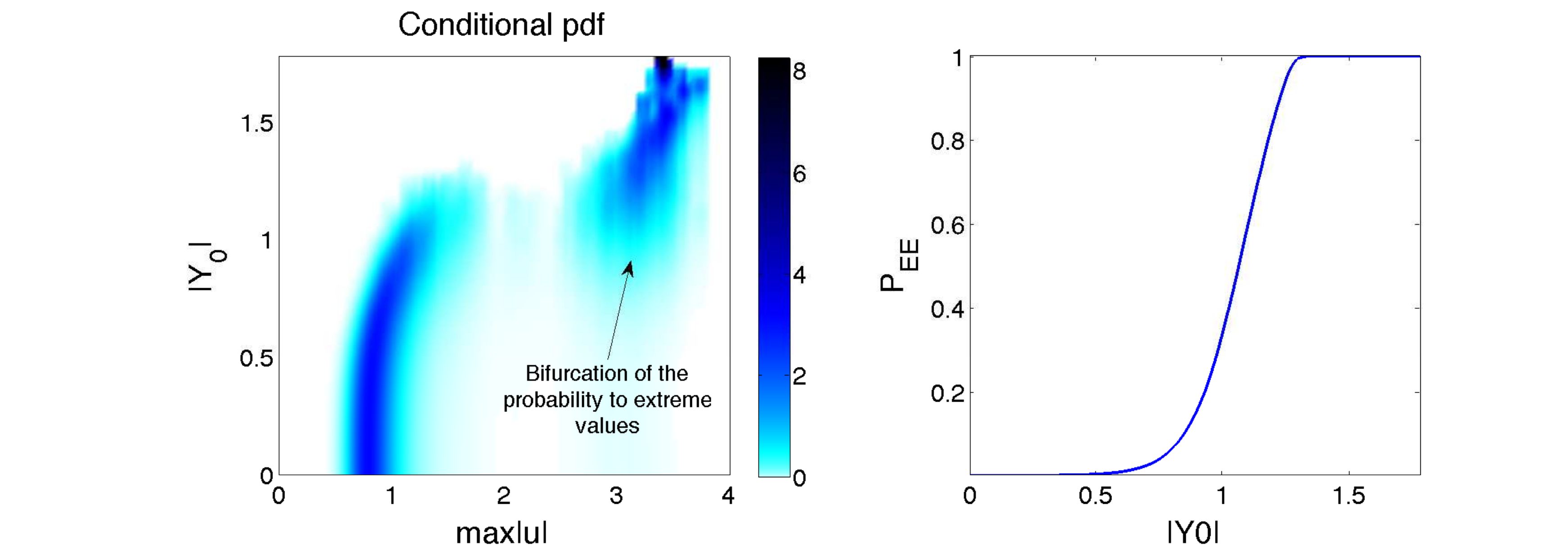}\caption{Left: Family of conditional densities of maximum nearby $|u|$ given
current value of $|Y_{0}|$. Right: Probability of an extreme event
in near future given $|Y_{0}|$.}

\label{fig:Y0CondPDF} 
\end{figure}

In Figure~\ref{fig:Y0CondPDF}, we display the family of conditional
density functions corresponding to (\ref{eq:Y0CondPDF}). Clearly
there is a critical value of $|Y_{0}|$ which, when exceeded, implies
that a nearby extreme event is highly likely. We now compute the probability
of a nearby extreme event, given a particular value of $|Y_{0}|$.
Here we define an extreme event as an instance where $|u|>2.5$. This
value is greater than twice the significant wave height--here taken
to be four times the typical deviation of the wave field (and is consistent
with the informal definition of rogue waves in the ocean \citep{Dysthe08}).
This probability, displayed in the right half of Figure~\ref{fig:Y0CondPDF},
is simply $F_{\mathcal{Y}_{0}}(2.5)$ from (\ref{eq:Y0CondPDF}).

There is a clear bifurcation in the distributions displayed in Figure~\ref{fig:Y0CondPDF}.
When $|Y_{0}|>1.1$, the likelihood of an upcoming extreme event increases
dramatically. We may use the definition of $||v_{0}||$ to compute
the energy level at which this bifurcation occurs at the critical length
scale $L=0.031$: 
\begin{align*}
1.1||v_{0}|| & =1.1\sqrt{\int e^{-2(x/L)^{2}}\, dx}=1.1\sqrt{L}\sqrt[4]{\pi/2}\approx0.22
\end{align*}
This critical energy value of 0.22 agrees well with our results from
Section~\ref{sec:energyLocalization}, where we found that Gaussian
initial data localized at length scale $L=0.031$ initiated extreme
events if the energy level exceeds approximately 0.2 (see Figure~\ref{fig:MMTCritEnergy}).
These results from Section~\ref{sec:energyLocalization}
were performed for localized states with a zero background, but the
agreement of this analysis with the bifurcation energy level in Figure~\ref{fig:Y0CondPDF}
is significant. Specifically, it suggests that the same localized
energy instability discussed in Section~\ref{sec:energyLocalization}
for toy examples triggers the formation of extreme events out of more
complex states.

We now analyze the performance of the computed extreme event probability
data from Figure~\ref{fig:Y0CondPDF} as a predictive scheme. At
a given time, we compute $|Y_{0}(x_{c},t)|$ for various values of
$x_{c}$ and use our compiled statistics (Figure~\ref{fig:Y0CondPDF})
to estimate the probability of an extreme event. If this probability
exceeds 0.8, we predict that an extreme event will occur. Choosing
a larger probability threshold value would decrease our rate of false
positives; choosing a smaller value would increase this rate but would
have the benefit of increasing the amount of time by which extreme
events are predicted in advance. We found that a probability threshold
of 0.8 provides a reasonable balance between these two effects (false
positive rate versus advanced warning time). Essentially the predictive
scheme measures the probability that a given combination of phases
between wave components (the current form of the wave field) belongs
to the domain of attraction of an extreme wave.

We tested this scheme on 50 simulations of (\ref{eq:MMT}). These
simulations were \emph{not} used to compute the statistics in (\ref{eq:Y0CondPDF})
and Figure~\ref{fig:Y0CondPDF}. In these simulations, we predicted
an extreme event 191 times, and 155 correctly predicted an extreme
event, meaning that the false positive rate was only 18.9\%. There
was only 1 extreme event that was not predicted by our scheme, which
means that the false negative rate was less than 1\%. As mentioned
in Section~\ref{sec:gaborStats}, in addition to preceding extreme
events, large values of $Y_{0}$ can occur after extreme events as
energy is being transferred to larger scales. However, large values
of $|Y_{0}|$ in this particular situation do not actually imply that
an extreme event is forthcoming. To avoid false positive predictions
that such behavior would generate, we ``turn off'' our predictive
scheme in the spatial region nearby the extreme event for the following
1 time unit.

\begin{figure}[ptb]
\centering \includegraphics[width=5.5625in,height=2.1871in]{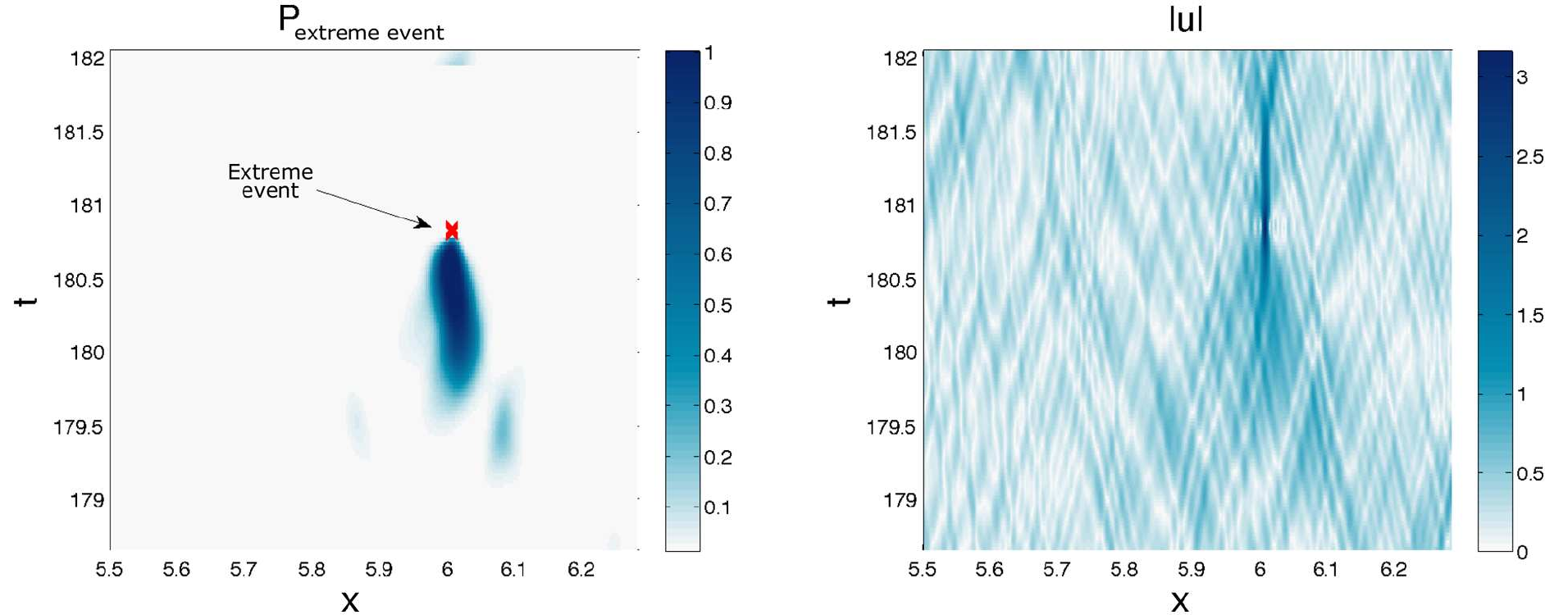}\caption{Left: Spatial distribution of probability for a nearby extreme event.
Right: $|u|$ as a function of space and time. The above figure shows
the spatial skill of our predictive scheme}

\label{fig:predSpace} 
\end{figure}

\begin{figure}[ptb]
\centering \includegraphics[width=5.5824in,height=2.6852in]{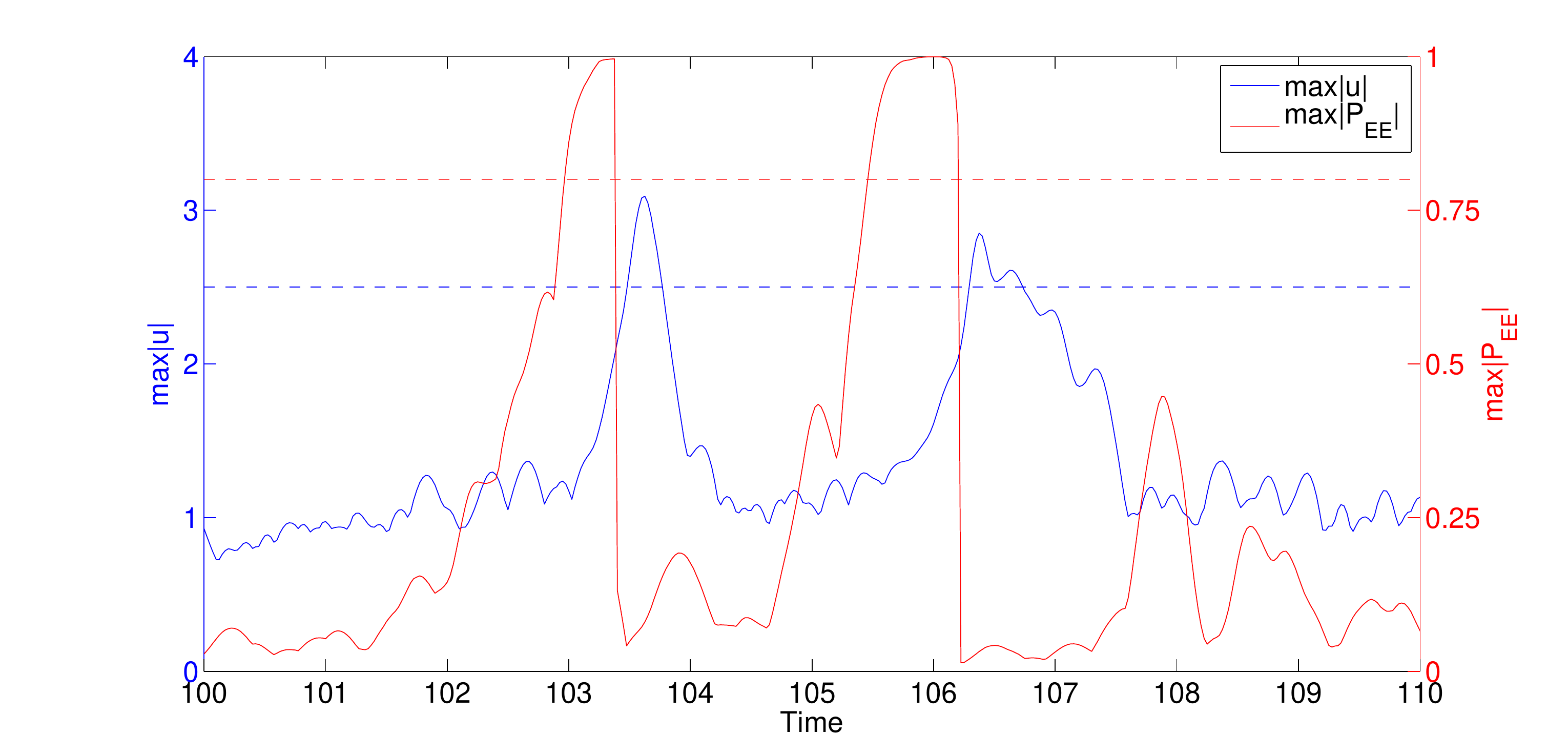}\caption{Spatial maximum values of $|u|$ and extreme event probability, showing
that our predictive scheme is able to give advance warning of extreme
events.}

\label{fig:predTime} 
\end{figure}

In Figure \ref{fig:predSpace} we present the spatial distribution
of the probabilistic predictor (left) and the actual wave field (right)
for one random realization. As we observe the computed criterion captures
accurately not only the temporal but also the spatial position of
the extreme wave. The same conclusions can be drawn from Figure \ref{fig:predTime}
where our prediction scheme is often able to predict extreme events
a full 1-2 time units in advance . 

\section{Discussion and Conclusions}

We have examined the synergistic activity of nonlinearity, dispersion,
and dissipation towards the formation of extreme events in a one-dimensional
prototype system that possesses four-wave resonant interactions, the
focusing MMT equation. The latter provides a relatively simple model
of extreme events arising out of a nearly Gaussian background with
broad-band spectrum, mimicking in this way many features of rogue
waves in the ocean. Through analytical and numerical tools we have
shown that the MMT is highly sensitive to \emph{localized} perturbations
of a particular critical length scale (Figure~\ref{fig:MMTCritEnergy}),
which we analyze thoroughly. We show the existence of a family of
solutions with a scale-invariance property and based on this fact
we quantify the required localized amount of energy that triggers
an extreme event. Although the existence of a critical energy level
for extreme events is certainly related to the modulation instability,
our analysis illustrates that even zero-backroung-energy states can
lead to an extreme event if a localized perturbation of appropriate
lengthscale and intensity is applied. These localized perturbations
can occur randomly through the dispersive propagation of waves that
have with random relative phase. 

We have illustrated that these extreme events are characterized by
low-dimensionality and we have use a spatially localized basis, a
Gabor basis to describe their characteristics. By performing a statistical
analysis of the Gabor coefficients we have been able to develop an
inexpensive predictive scheme that is reliable with few false positives
and false negatives. Furthermore, our scheme shows a high degree of
spatial skill and issues warnings in advance (often 1-2 time units
before the extreme event). Future research efforts include the extension
of the prediction window by combining the presented approach with
nonlinear filtering techniques \citep{Majda_filter}. We are also
interested on applying the presented framework in more realistic two
dimensional nonlinear wave models and a current research effort is
focused on the wave equation by Trulsen et. al. \citep{trulsen2000}
which, like the MMT model, includes the exact dispersion relation
for gravity waves over deep water.

\textbf{Acknowledgments.} This research effort is funded by the Naval
Engineering Education Center through grant 3002883706; We are grateful
to Dr. Craig Merrill (technical point of contact for this project)
for numerous motivating discussions and support. We would like to
thank Prof. Andrew Majda who suggested the MMT model as a prototype
system for extreme events as well as for numerous other stimulating
comments. We are grateful to Dr. Ian Grooms for providing a version
of his numerical solver for the MMT system, as well as to Lake Bookman
for helpful discussions.

\bibliographystyle{ieeetr}
\bibliography{main}

\end{document}